\documentclass{article}
\usepackage[utf8]{inputenc}
\usepackage{graphicx}
\usepackage{amsmath}
\usepackage{amssymb}
\usepackage[T1]{fontenc}
\usepackage{authblk}
\usepackage{xcolor}
\usepackage{soul}

\title{Towards 3D-Printed Inverse-Designed Metaoptics}
\author[1]{Charles~Roques-Carmes\thanks{chrc@mit.edu}}
\author[2]{Zin~Lin}
\author[3,4]{Rasmus~E.~Christiansen}
\author[1,5]{Yannick~Salamin}
\author[6]{Steven~E.~Kooi}
\author[5,6]{John~D.~Joannopoulos}
\author[2,5]{Steven~G.~Johnson}
\author[1,5]{Marin~Solja\v{c}i\'{c}}
\affil[1]{Research Lab of Electronics, Massachusetts Institute of Technology, Cambridge MA 02138, USA}
\affil[2]{Department of Mathematics, Massachusetts Institute of Technology, Cambridge MA 02138, USA}
\affil[3]{Department of Mechanical Engineering, Technical University of Denmark, Nils Koppels All\'{e}, Building 404, 2800 Kongens Lyngby, Denmark}
\affil[4]{NanoPhoton---Center for Nanophotonics, Technical University of Denmark, {\O}rsteds Plads 345A, DK-2800 Kgs. Lyngby, Denmark.}
\affil[5]{Department of Physics, Massachusetts Institute of Technology, Cambridge MA 02138, USA}
\affil[6]{Institute for Soldier Nanotechnologies, 500 Technology Square, Cambridge, MA}

\usepackage[isbn=false,url=false,style=nature]{biblatex}    

\begin{document}

\maketitle

\begin{abstract}
Optical metasurfaces have been heralded as the platform to integrate multiple functionalities in a compact form-factor, potentially replacing bulky components. A central stepping stone towards realizing this promise is the demonstration of multifunctionality under several constraints (e.g. at multiple incident wavelengths and/or angles) in a single device --- an achievement being hampered by design limitations inherent to single-layer planar geometries. Here, we propose a general framework for the inverse design of volumetric 3D metaoptics via topology optimization, showing that even few-wavelength thick devices can achieve high-efficiency multifunctionality. We embody our framework in multiple closely-spaced patterned layers of a low-index polymer. We experimentally demonstrate our approach with an inverse-designed 3d-printed light concentrator working at five different non-paraxial angles of incidence. Our framework paves the way towards realizing multifunctional ultra-compact 3D nanophotonic devices.
\end{abstract}

\section{Introduction}

The design and realization of three-dimensional metastructures is the next stepping stone of nanophotonics, providing the basis for a new disruptive class of multifunctional photonic devices. The main promise of such devices is to realize multifunctional optics in a compact (wavelength scale) form-factor and replacing their bulky counterparts~\cite{Khorasaninejad2016MetalensesImaging}, with applications in microscopy, imaging, augmented and virtual reality~\cite{Khorasaninejad2016MetalensesImaging, Khorasaninejad2017Metalenses:Components, Li2021Meta-opticsReality}. While metasurfaces, single-layer structured surfaces at the nanoscale, have already demonstrated novel functionalities in the polarization-state control and conversion~\cite{Arbabi2015DielectricTransmission}, vortex beam generation~\cite{Devlin2017ArbitraryLight}, imaging~\cite{Khorasaninejad2017Metalenses:Components}, and spectral, or angular multifunctionalities~\cite{Shi2018Single-LayerFunctions}, their 2D configuration imposed by traditional lithography methods severely restricts multi-functional designs. It therefore becomes  critical to develop 3D nanophotonics platforms.   

In this work, we provide the first experimental demonstration of topology-optimized 3D-printed multifunctional metaoptics. Our work is based on a general framework in which large-scale structures are optimized to achieve simultaneous multifunctional operation while enforcing material and geometrical constraints. To exemplify the promise of this workflow, we have designed (through topology optimization), fabricated (via two-photon polymerization (TPP) in Nanoscribe IP-Dip), and experimentally demonstrated a 3D optical concentrator which focuses 5 different angles to the same focal spot. We show that the inverse-designed fabricated structure matches well with the theoretical performance, and propose avenues to scale our technique to thicker (more layers), high-efficiency, 3D-printable volumetric optics. We also show theoretically how few-layer metaoptics can achieve Strehl ratios (SR) well above 0.8 for a broad range of wavelengths, outperforming existing metasurfaces capabilities. 

Until now, the realization of multiple complex functionalities in a single metasurface has been thwarted by fundamental limitations of single-layer structures~\cite{Shim2020MaximalWaves}. While advanced (inverse-)design techniques (e.g. density-based topology optimization) have generated designs achieving optimized performance for single-layer structures~\cite{Pestourie2018InverseMetasurfacesb, Jiang2019Simulator-basedMetasurfaces,Sell2017PeriodicFunctionalities}, it is commonly acknowledged that 3D structures, for instance embodied in a compact (wavelength-scale) arrangement of multiple layers of a dielectric material, would enable a quantum leap in terms of multifunctionality, while conserving the main value proposition of metasurfaces: compactness~\cite{Lin2018Topology-OptimizedMetaoptics, Lin2019OverlappingMetasurfaces, Lin2021ComputationalAberration, Rodriguez2018InverseNanophotonics, Mansouree2020MultifunctionalOptimization, Molesky2020Sections, Naseri2021AMetasurfaces,Shim2020MaximalWaves,Christiansen2020FullwaveMetalenses}. Despite this observation, there has previously been very little experimental work dedicated to compact multi-layer structures (which we will refer to as volumetric optics). Earlier work~\cite{Mansouree2020MultifunctionalOptimization} focused on adjoint-optimization of a 2.5D CMOS platform to realize a double-wavelength focusing metastructure. While 2.5D stacking of CMOS layers is a promising platform for the next generation of integrated photonic circuits, it still suffers from major challenges in layer-to-layer alignment and non-uniform performance, severely limiting the practically achievable number of layers and possible designs geometries~\cite{BenYoo2016HeterogeneousMicrosystems}.

Another factor hampering the realization of wavelength-scale volumetric optics has been the availability of three-dimensional photonic fabrication platforms. The metasurface community has therefore largely favored single-layer high-index platforms (such as silicon~\cite{Arbabi2015DielectricTransmission} and titania~\cite{Khorasaninejad2016MetalensesImaging}), allowing for greater light confinement and induced phase response. However, recent developments in three-dimensional photonics~\cite{Soukoulis2011PastMetamaterials}, enabled e.g. by two-photon polymerization~\cite{Hahn2020RapidMetamaterials}, have opened a path to arbitrary free-form subwavelength photonic structures. Because of their coarser fabrication constraints (in comparison with latest-generation lithography platforms) and the relatively smaller refractive index of the polymerized materials, the use of advanced design techniques---such as topology optimization---is of paramount importance to fully exploit the 3D degrees of freedom. 

\begin{figure}
    \centering
    \hspace*{-0.5in}
    \includegraphics[scale=0.8]{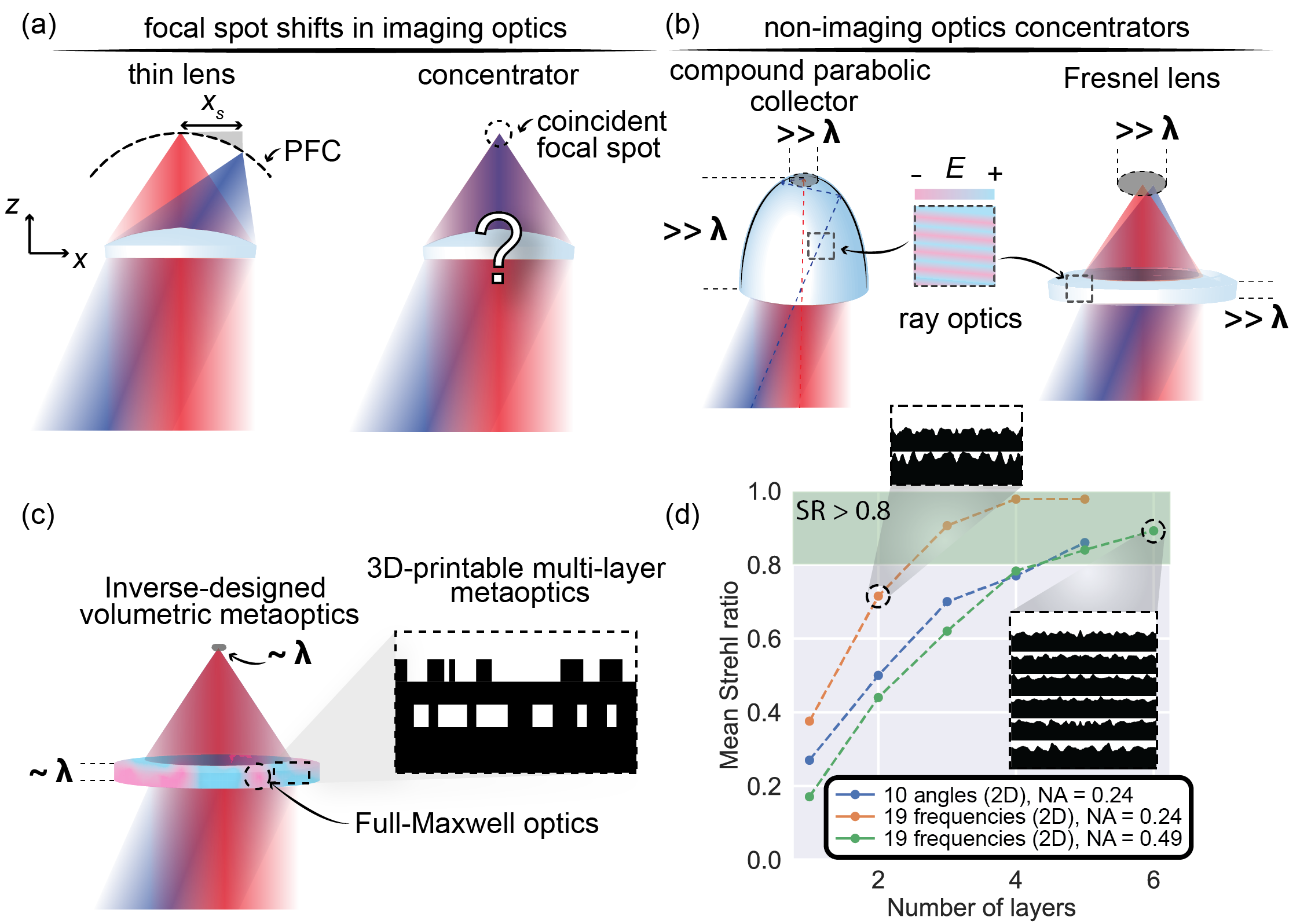}
    \caption{\textbf{Comparison of several approaches to light concentration.} (a) Left: A thin lens illumated by a plane wave at non-normal incidence focuses light at a position shifted in the $x$ and $z$ directions, compared to normal incidence illumination. This effect is known as the Petzval Field Curvature (PFC). Right: An ideal imaging concentrator would focus light from on- and off-normal incidence in a wavelength-scale (diffraction-limited) focal spot at the same location. However, its realization is forbidden using linear reciprocal ray optics (see Supplementary Information, Section I). (b) Non-imaging optics concentrators are typically used to concentrate light in a spot much larger than the wavelength, with reflective (left) and refractive (right) optical devices whose dimensions are much larger than the wavelength themselves, thus relying on ray or diffractive optics design techniques. (c) Our approach consists in inverse-designed volumetrics optics, focusing on- and off-normal incidence plane waves at the same, diffraction-limited, focal spot subject to the fundamental efficiency limits set by brightness theorems \cite{Hsu2019ScatteringWaves}. The device itself is a few wavelengths thick. Our approach is embodied in a 3D-printable multi-layer single-piece metaoptics, realized in a low-index polymer. (d) Performance scaling on the conventional imaging task of removing only the $z$-component of the PFC (different angles going to different $x$ locations, but same $z$ location).}
    \label{fig:fig1}
\end{figure}

\section{Framework}

In this section, we show that metaoptics --- which are inverse-designed for 3D-printing in low-index polymers --- offer a number of benefits over traditional geometric optics and metasurfaces. Our volumetric metaoptics are few-wavelength-thick non-intuitive structures that are optimized to simultaneously realize a set of optical functionalities. We consider metaoptics that are inverse-designed with density-based topology optimization~\cite{Bendse2004TopologyMaterial, Jensen2011TopologyNano-photonics}. This allows us to design structures ``pixel by pixel'' over large diameters ($\gg 100 \lambda$), with little \textit{a priori} knowledge of the final structure, while allowing for the enforcement of fabrication constraints. We exemplify our approach with 3D-printable closely-packed layers of a low-index polymer (Nanoscribe IP-Dip).

Our design technique can be summarized as follows: the design problem is recast as a constrained continuous optimization problem where the quantity to be optimized (the power at a given location, or Strehl ratio) is formulated as an objective function of the state (electromagnetic fields) and design (material distribution) variable fields in a given region of space. We discretize each of the device layers into subwavelength pixels, each of which may vary continuously between 0 and 1 (representing a continuous transition between the permittivity of the background material and IP-Dip). 

Once set up, the optimization problem is solved using a gradient-based optimization algorithm (e.g. the Method of Moving Asymptotes~\cite{Svanberg2002AApproximations}). Since the thickness of each layer is wavelength-scale, and multiple scattering events occur between layers, the full Maxwell equations must be used to model and optimize such structures (in contrast with cascaded non-interacting optical surfaces, a common technique in modern optical engineering~\cite{Smith2000}, which was recently transferred to metasurfaces~\cite{Groever2017Meta-lensRegion, Zhou2018MultilayerMetaoptics, Arbabi2016MiniatureAberrations, ZhouMultifunctionalMetasurfaces}). To ensure that a physically realizable design results from solving the optimization problem, a filter-and-project procedure is applied during the optimization process~\cite{Christiansen2021InverseTutorial}. Further information about the techniques we employ can be found in the Supplementary Material and in previous works~\cite{Lin2018Topology-OptimizedMetaoptics, Lin2019OverlappingMetasurfaces, Lin2021ComputationalAberration, Christiansen2020FullwaveMetalenses}.

We use topology optimization to design a cylindrical concentrator working at 1550 nm (i.e., a concentrator that focuses onto a line). Since perfectly concentrating light from multiple angles is prohibited by reciprocal linear optics, there is no intuitive design nor ideal phase profile to realize such a functionality. This is in contrast to achromatic lenses, axicons, holograms, orbital angular momentum and polarization-state converters, where a target phase profile can be analytically computed and locally fitted with a library of pre-computed unit cells. Beyond ray optics, there is still a fundamental trade-off between the focal spot intensity and spot size, even for concentrators working at a single angle of incidence (lenses) \cite{Shim2020MaximalWaves}. Therefore, the use of inverse-design techniques such as topology optimization is not just optional, but a requirement for the design of nanophotonic concentrators. Such nanophotonic concentrators could find applications in photovoltaics as compact solar concentrators~\cite{Polman2012PhotonicPhotovoltaics, Lin2018Topology-OptimizedMetaoptics}, in high resolution maskless photo-lithography~\cite{Menon2005MasklessLithography}, and as compact photonic multiplexers. Given its diffraction-limited performance and its compact form-factor, the proposed concentrator could significantly reduce the footprint of photovoltaic devices and allow the use of microscale photodetectors.

Volumetric metaoptics can offer solutions to inherent drawbacks of traditional optics and metasurfaces. For example, conventional thin lenses and their nanophotonic counterparts (metalenses) suffer from Petzval field curvature (PFC)~\cite{Smith2000}: when an off-normal-incidence plane wave impinges on the lens, the focus is shifted in the transverse $x$ and longitudinal $z$ directions compared to their location under normal-incidence illumination (see Figure~\ref{fig:fig1}(a)). There has been significant effort in optical engineering to eliminate the PFC, such as the design of flat-field (plan) multi-lens objectives~\cite{Smith2000}. Such objectives are typically composed of multiple thin lenses whose Petzval sum cancels out (and also corrects other aberrations, such as chromatic aberrations: in this case called plan achromats). While correcting the $z$-component of the PFC is achievable with intuitive optical designs, the task of concentrating light incident from multiple off-normal angles at the same (diffraction-limited) focal spot, as illustrated in Figure~\ref{fig:fig1}(a, right), is even more challenging, and cannot be addressed with conventional single-layer thin-lens design techniques. It is even prohibited in reciprocal linear optics, as reviewed in the Supplementary Information (SI, Section I). In the field of solar concentrators for photovoltaic applications, concentration is typically realized with non-imaging (reflective or refractive) optics, for a narrow angular bandwidth at the cost of a focal spot size much larger than the wavelength (and therefore far from diffraction-limited performance)~\cite{Khamooshi2014AConcentrators}. Two popular examples of such refractive and reflective optical concentrators are shown in Figure~\ref{fig:fig1}(b). Furthermore, such non-imaging concentrators operate in the ray or diffractive optics limit, where light ray trajectories are shaped by features much larger than the wavelength, thus resulting in bulky devices. In the following paragraphs, we show that topology optimization can address this challenging design problem, while enforcing fabrication constraints, allowing for its experimental realization. 

In addition to concentrators, we also study the theoretical performance of volumetric metaoptics on the more conventional task of plan-achromatic focusing, characterized by SR as a function of the number of layers. SR is defined as the ratio of the focal peak of the designed lens to the ideal Airy peak normalized by the transmitted power (see Supplementary information). As shown in Figure~\ref{fig:fig1}(d), high-performance focusing (SR $\geq$ 0.8, a gold standard in optical engineering~\cite{Smith2000}) can be achieved over several angles or several frequencies with only a few more layers than what we experimentally demonstrate in this work. Recently, we proposed several topology optimization techniques to discover new multi-layer designs made of low-index polymers with state-of-the-art performance~\cite{Lin2021ComputationalAberration, Christiansen2020FullwaveMetalenses}. 

\begin{figure}
    \begin{center}
    \hspace*{-1.0in}
    \includegraphics[scale=0.85]{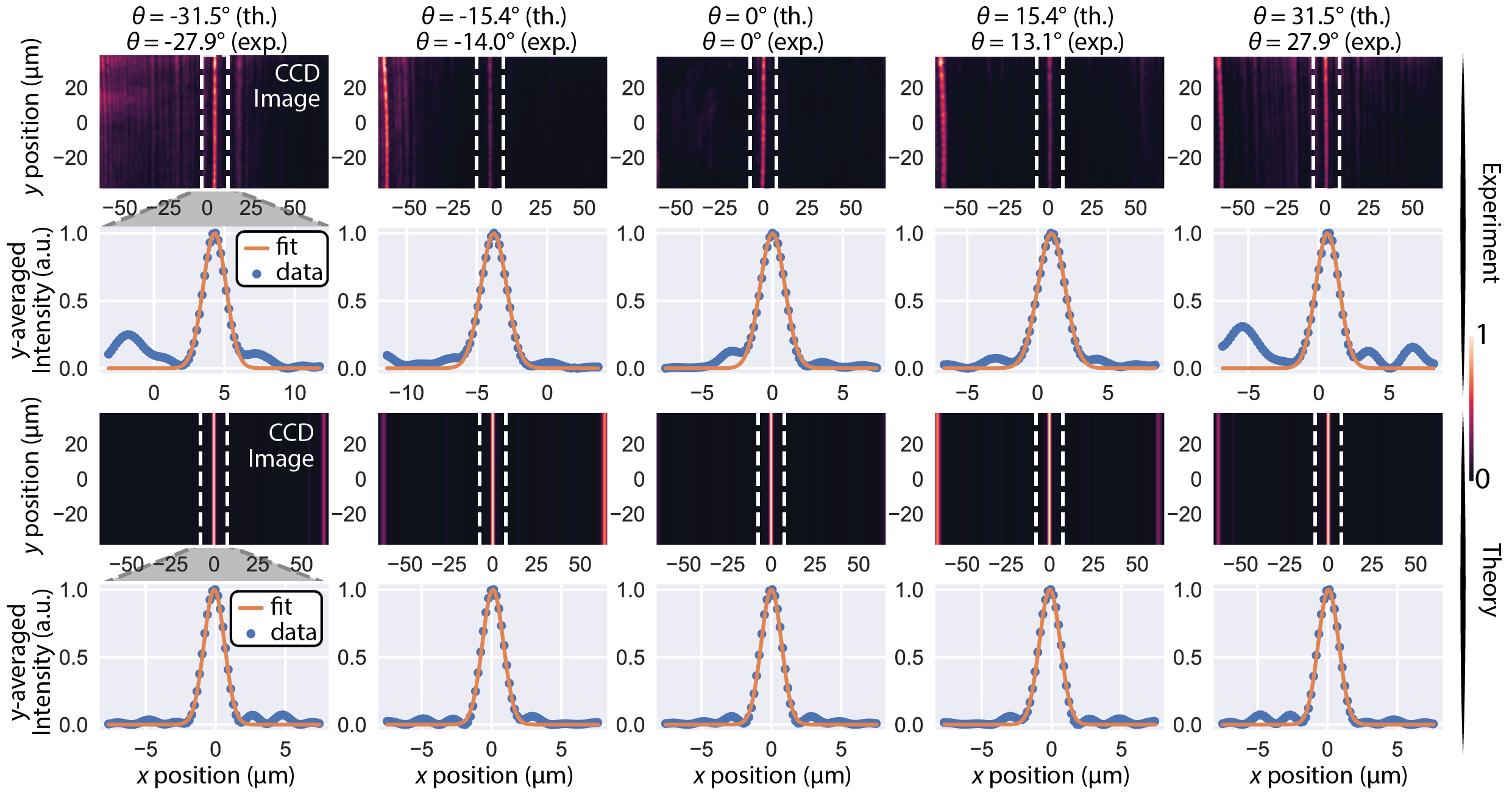}
    \caption{\textbf{Light concentration with inverse-designed 3D-printed multi-layer metaoptics.} Columns correspond to various measured and simulated angles (whose values is indicated at the top of the column). The first two rows show experimental results. The last two rows show numerical simulations. In the rows showing the CDD images, the dashed white square corresponds to the area over which the signal is averaged to give the row below. In this figure each plot is normalized independently.}
    \label{fig:fig2}
    \end{center}
\end{figure}

\section{Experimental Results}

In this section, we experimentally demonstrate a two-layer cylindrical concentrator. The full design (spatial permittivity distribution obtained via topology optimization) is shown in the Supplementary Information. To account for fabrication constraint in TPP, the minimum feature size in the design was fixed to 310~nm (after applying filters---the pixel size, used to discretize Maxwell's Equations, being $0.02\lambda$) in the transverse $x$ direction and 930nm in the longitudinal $z$ direction. The design is optimized to maximize intensity at the same focal spot [located at $(x,z) = (0, f)$ with $f = 232~\mu$m] for five incident angles in air ($\theta_\text{th} \in \lbrace -31.5^\circ, -15.4^\circ, 0, 15.4^\circ, 31.5^\circ \rbrace$). 

The experimentally measured farfield focal lines in the plane $z=f$ are shown in Figure~\ref{fig:fig2} (first row) for the measured angles corresponding to a zero-shift focal line $\theta_\text{exp} \in \lbrace -27.9^\circ, -14.0^\circ, 0, 13.1^\circ, 27.9^\circ \rbrace$ (a relative error of 11.7\% on average, compared to the design values). For comparison, the theoretical focal plane images and corresponding focal lines at the design angles are shown in Figure~\ref{fig:fig2} (bottom two rows). The $y$-averaged intensity, also shown in Figure~\ref{fig:fig2} is a peak whose full-width at half-maximum (FWHM) is close to the diffraction limit for all measured angles $\text{FWHM}_\text{exp} \in \lbrace 1.92, 2.15, 2.08, 2.28, 1.97 \rbrace \pm 0.19~\mu$m (a relative increase of 20\% compared to the designed values, which only differ by 1-3\% from the diffraction limited behavior of $0.44\lambda/\text{NA} = 1.71~\mu$m for a cylindrical concentrator working at a wavelength of $\lambda = 1.55~\mu$m and a numerical aperture of $\text{NA} = 0.44$). 

\begin{figure}
    \begin{center}
    \hspace*{-1.0in}
    \includegraphics[scale=0.85]{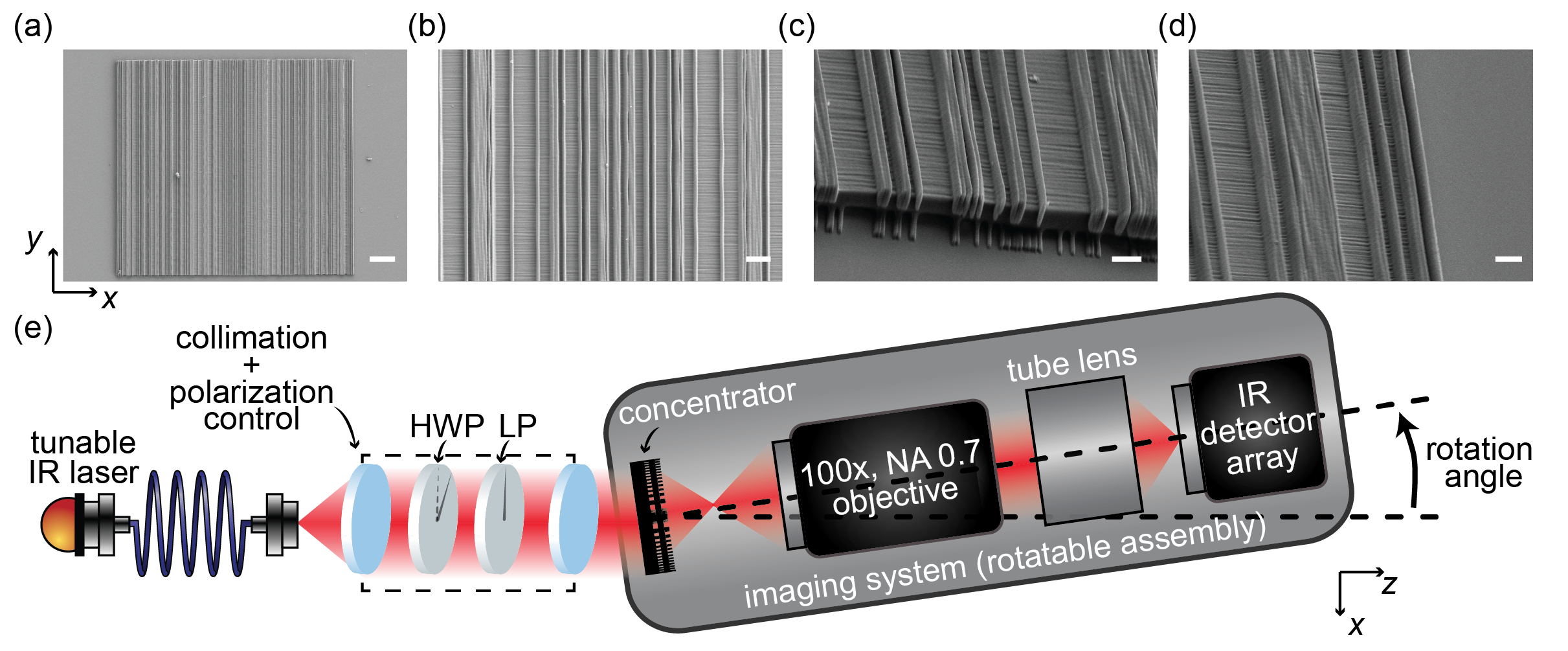}
    \caption{\textbf{Sample and experimental setup.} (a-d) Scanning Electron Micrographs (SEM) of the measured sample. (a): Full view. Scale bars: 20 $\mu$m (a), 3 $\mu$m (b), 2 $\mu$m (c), 2 $\mu$m (d). (e) Schematic of the imaging experimental setup.}
    \label{fig:fig3}
    \end{center}
\end{figure}

Essential to our approach is the use of TPP to fabricate 3D structures. One of the main advantages of TPP compared to multi-layer CMOS or wafer bonding is the scalability to multiple layers of patterned material in a single fabrication run, thus removing the need for wafer-to-wafer alignment as well as the necessity of chemical-mechanical planarization~\cite{Zantye2004ChemicalApplications}. Since TPP is limited to coarser feature sizes (down to $\sim 200$ nm) and forbids the existence of closed cavity in the final structure (where unpolymerized resist cannot be removed during development), we choose to embody our framework in a multi-layered arrangement of patterned/unpatterned extruded ridges 930 nm tall and 310 nm wide. Density and binarization filters~\cite{Jensen2011TopologyNano-photonics} are applied in the design optimization to ensure the final design does not violate these fabrication constraints. A discussion of different paths to realizing full 3D inverse-designed and fabricated structures with rotation-invariant geometries is included in the next section of this article and in the SI.

Scanning electron micrograph (SEM) images of the measured sample are shown in Figure~\ref{fig:fig3}(a-d). The total size of the sample is $200~\mu$m along the $x$ direction and $180~\mu$m along the $y$ (extruded) direction (this way the full structure remains within one field of view of the TPP printer). Top-view SEMs show the top layer made of straight polymer lines with minimum feature size $310$~nm extruded along the $y$-direction. The bottom layer, hidden under an unpatterned layer of 930~nm thickness, is apparent in tilted SEM images (see Figure~\ref{fig:fig3}(c)).
To measure the concentrator's performance at various angles of incidence, we mount the sample at the center of a heavy-load rotating stage, as shown in Figure~\ref{fig:fig3}(e). The scattered farfield, when exciting the concentrator with a plane wave incident at angle~$\theta$, is measured by rotating the stage by~$\theta$, while keeping the illumination optics fixed. The imaging assembly position with respect to the concentrator remains fixed, thus allowing us to measure the concentrator's response at various angles of incidence without moving the sample. Since the sample is concentrating most of the light close to the center of the field of view at various off-normal angles, the focal lines are for the design angles in the detector field of view (of about $\sim 100~\mu$m in the $x$ direction). 

\begin{figure}
    \begin{center}
    \hspace*{-1.0in}
    \includegraphics[scale=0.85]{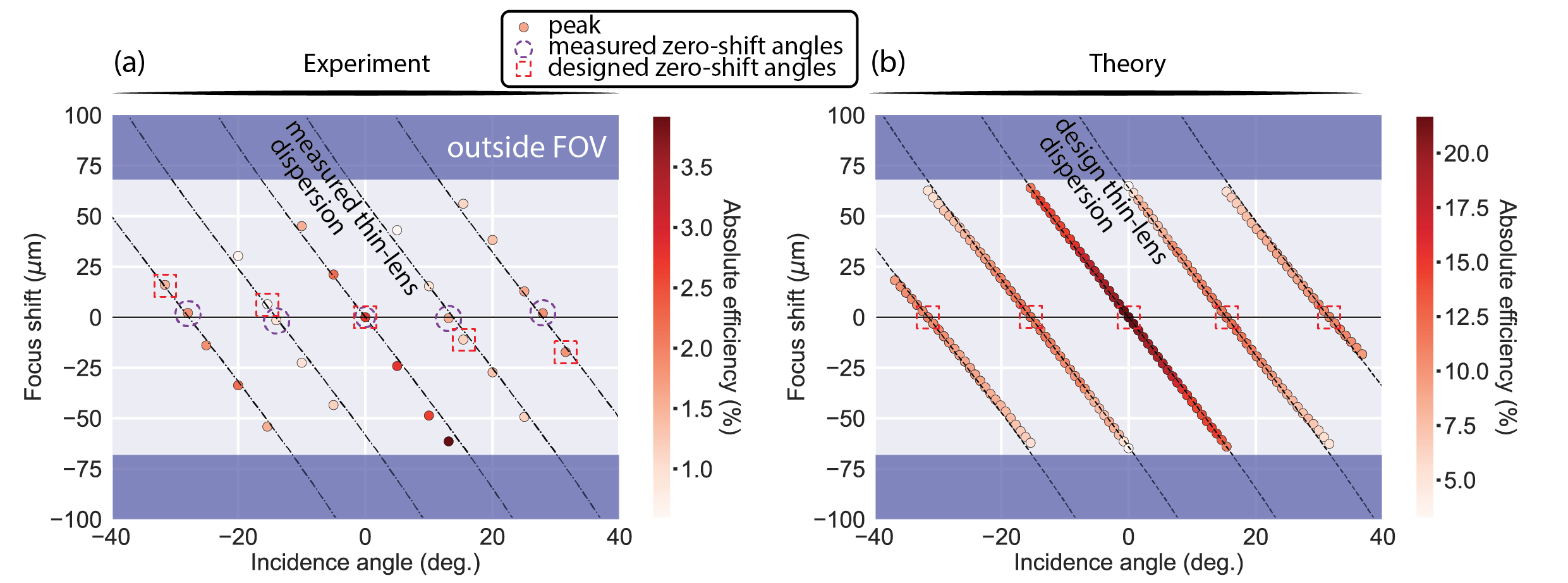}
    \caption{\textbf{Dispersion engineering.} (a) Experimental results. (b) Numerical simulations. The blue shaded areas show positions outside the camera field of view (FOV). Each dot shows the position of a focal line at a given incidence angle. The color of the dot indicates the absolute efficiency (see definition in the Methods). The dashed circles and squares show the measured and designed zero-shift angles, respectively. The dashed black lines show the thin lens dispersion centered at the measured (a) or designed (b) zero-shift angles.}
    \label{fig:fig4}
    \end{center}
\end{figure}

Inverse design is a particularly attractive technique for tasks that do not have intuitive physical solutions, but also for multi-constrained tasks \cite{Piggott2015InverseDemultiplexer, Dory2019Inverse-designedPhotonics, Yang2021Inverse-designedTransmitters}, such as plan-achromatic focusing or multi-angle concentration of light. While a human might know how to design a close-to-optimal structure for a single constraint, a gradient-based numerical optimization method has the ability to systematically and efficiently traverse a vast design space and identify an optimized design satisfying multiple constraints at the same time. However---unless imposed as a constraint in the design process---the resulting device may lack immediate physical interpretability, as is the case for the nanophotonic concentrator. In order to obtain a better understanding of the physical mechanisms responsible for the inverse-designed concentrator's behavior, we perform a series of additional measurements as detailed next.

We measure the concentrator's response at various angles within the incident angular range $\theta \in [-31.5^\circ, 31.5^\circ]$, with results shown in Figure~\ref{fig:fig4}(a). Each dot corresponds to a focal line detected in the detector's field of view, with the dot color's proportional to the estimated absolute efficiency of the corresponding focal line. Our experimental methods used to measure the focus $x$-shift and the efficiencies are reported in the Methods section. We first observe that the experimental measured dispersion matches closely that of the theoretical design shown in Figure~\ref{fig:fig4}(b), (the observed mismatch in efficiency is discussed below). The dispersion allows us to better understand the design blueprint identified by our inverse-design algorithm: namely, multiple replicas of the thin-lens dispersion $x = f\tan\theta$ shifted by the design angles. Even though we observed a discrepancy in the measured zero-shift angles compared to the values targeted in the design process, this observation remains true (but the value by which the thin-lens dispersion replica shifts is that of the measured zero-shift angle, instead of the designed zero-shift angle). In Figure~\ref{fig:fig4}(b), each thin-lens dispersion shifted replica is more efficient close to its zero-shift angle. Based on these observations, the optimized design can be interpreted as a ``superposition'' of several optimized designs working for a single constraint, and the resulting optimized device switches between each single-constraint operation when close to its corresponding design angle. 

\section{Discussion}

The experimental demonstration presented in this work pushes the feature size of TPP fabrication close to its minimal values $\sim 200$--300~nm using a commercially-available two-photon polymerization machine (Nanoscribe). Current and future developments in resist chemistry and TPP may relax these fabrication constraints further~\cite{Hahn2020RapidMetamaterials}. For the time being, our framework could be applied directly to longer-wavelength operation, where commercially-available resists and tools are transparent and can be patterned in deeply subwavelength patterns (for instance, the mid-infrared regime of $\sim 4~\mu$m seems ideal for IP-Dip~\cite{Fullager2017InfraredCm1}).

The measured efficiencies reported in Figure~\ref{fig:fig4}(a) are about a factor of 5 smaller than predicted by the numerical results, a mismatch much larger than the observed discrepancy on the other figures of merit reported in this paper (namely the FWHM, dispersion, and measured focal distance). While a mismatch in refractive index of polymerized IP-Dip~\cite{Li2019UVEllipsometry, Dottermusch2019Exposure-dependentLayers} could account for an error of about $\sim 1\%$ in performance, the SEM shown in Figure~\ref{fig:fig3}(a--d) suggests that surface roughness and fabrication inaccuracies could explain the larger mismatch in efficiency. Additional SEMs are shown in the Supplementary Information, where we show that a few parallel ridges collapse on each other, most likely due to capillary forces (during unpolymerized resist removal or measurement in a humid environment). This could be avoided by use of critical-point drying techniques and/or measurement in a dry environment. Other factors, such as non-uniformity of ridge widths and heights, and mismatch with the theoretical design, may further explain the efficiency gap. These observations suggest utilizing encapsulation of such devices as a technique to mitigate environmental damage, or limiting their use to dry environments. 

We demonstrated volumetric optics limited to multi-layer metasurfaces, an approach that has been favored in the metasurface inverse-design community for its ease of parametrization and mechanical stability. However, this approach prevents the fabrication of axisymmetric structures~\cite{Christiansen2020FullwaveMetalenses} without additional constraints. To realize full 3D structures, one can resort to a few approaches. First, adding intermediate air layers that allow unpolymerized polymer to be extracted after printing. Then, additional mechanical constraints and evacuation channels need to be implemented for unpolymerized resist removal. We propose one such schematic design in the Supplementary Information. Second, recent 3D nanofabrication techniques have been proposed, where the printed structure creates an index contrast trapped into a polymer matrix. While current techniques are limited to metallic structures~\cite{Oran20183DScaffolds}, dielectric nanoparticles could be patterned into 3D structures in the future with similar techniques. Third, more general degrees of freedom could be considered with topology optimization, while enforcing fabricability constraints. There has been recent work aimed at enforcing structural integrity while achieving similar performance as unconstrained designs~\cite{Augenstein2020InverseIntegrity}.

In summary, we have presented a general framework for inverse-designing and fabricating volumetric metaoptics to realize optimized and compact multifunctional optical devices. Our framework is experimentally demonstrated using multi-layered nanopatterned structures fabricated via TPP in a low-index polymer. We experimentally demonstrate a two-layer optical concentrator, focusing incident light at five different angles into the same wavelength-scale spot. Our results pave the way toward the realization of 3D-printed inverse-designed highly-multifunctional metaoptics. 

\section{Methods}

\subsection{Experimental setup}
A schematic of the experimental setup is shown in Figure~\ref{fig:fig3}. A tunable continuous wave infrared laser input is fiber-coupled. The fiber coupling output is collimated by two lenses, and the incident polarization is controlled by a half-wave plate (HWP) and a linear polarizer (LP). The concentrator is positioned at the center of rotation of a heavy-load motorized rotating stage. The stage supports the concentrator and a fixed imaging system assembly (100X IR Mitutoyo objective, $f = 200$ mm tube lens and an infrared detector) which rotates with the concentrator. When the stage rotates, the angle of incidence of the collimated beam rotates, but the angle between the concentrator plane and the imaging assembly is fixed, and the concentrator remains at the center of the objective field of view.

\subsection{Focal line and shift measurement}
The $x$ position of focal lines is calculated by taking $y$-averages of the measured CCD images. For each measured angle, we measure three focal line positions: $x_0$, $x_\theta$, and $x_0'$, corresponding to the peak position in the farfield measured at normal incidence, at the angle of interest, and back at the normal incidence (the stage is rotated twice, in the $0\rightarrow \theta$ direction first, then in the $\theta \rightarrow 0$ direction). The value of the $x$-shift plotted in Figure~\ref{fig:fig4}(a) are defined as $\Delta x \pm \delta(\Delta x)$ where $\Delta x = \frac{1}{2}(2 x_\theta - x_0 - x_0')$ and $\delta(\Delta x) = |x_0 - x_0'|$. Therefore, the value of the shift is measured with respect to the focal line at normal incidence before and after rotating the stage and the relative error is defined as the distance between the focal line at normal incidence before and after rotating the stage (if these two focal lines coincide exactly, then the measurement is exact). 

\subsection{Efficiency estimate and power calibration}
From the focal lines, we estimate the value of the FWHM from fitting the peak to a Gaussian function with standard deviation $\sigma$: $\text {FWHM} =2\sqrt{2\ln 2}\ \sigma\approx 2.35482\sigma$. The absolute efficiency (encoded in the color of the dots of Figure~\ref{fig:fig4}) is defined as the measured power in $\pm 3~\text{FWHM}$, accounting for optical losses in the setup: $\eta = \frac{P_{\pm3~\text{FWHM}}}{P_\text{in}\eta_\text{opt}}$. $\eta_\text{opt}$ accounts for the following factors: (1) Reflection and losses through the objective, tube lens, and sample substrate; (2) Normalization of the input power to the sample area (the incident beam is larger than the sample to ensure a quasi-uniform illumination with a measured diameter of $\sim750~\mu$m).

The power measured by the detector is calibrated by measuring the detector's response with a beam of known intensity. As indicated by the detector provider, it is characterized by a non-linear power law with exponent $\sim0.7$. More details and calibration measurements are provided in the Supplementary Information. 

\subsection{Sample fabrication}
The metaoptics concentrator was fabricated using a commercial TPP system (Nanoscribe Photonic Professional GT) on a $700$-micron-thick fused silica substrate. For this purpose, piezo actuators move the sample in the out-of-plane direction after fabricating each layer. Geometrical parameters and dose (scanning speed and laser power) are optimized with a dose test on this specific machine. In the in-plane direction the laser beam is guided by galvanometric mirrors parallel to the substrate. The writing of the patterned layer is performed in lines along the $y$ direction, while the writing in the unpatterned layer is performed in lines along the $x$ direction.

\section{Author contributions}
C.~R.-C. and Z.~L. conceived the general idea of this project. C.~R.-C., Y.~S., and S.~E.~K. performed the experiment. Z.~L. and R.~E.~C. developed the numerical methods. J.~D.~J., S.~G.~J., and M.~S. supervised the project.  C.~R.-C. wrote the manuscript with inputs from all authors. 

\section{Data availability statement}
The data that support the plots within this paper and other findings of this study are available from the corresponding authors upon reasonable request.

\section{Code availability statement}
The code that supports the plots within this paper and other findings of this study is publicly available at the following address: https://github.com/zlin-opt/meta2d\_inverse\_design\_oda.

\section{Acknowledgements}

The authors would like to thank Peixun Fan (University of Nebraska) for fabricating the sample and Wei-Ting Chen (AMS Sensors USA) for useful discussion on applications of concentrators. The authors would like to thank Jamison Sloan for critical reading and commenting of the manuscript. This work was partially funded by the U.S.~Army Research Office through the Institute for Soldier Nanotechnologies (award no.~W911NF-18-2-0048). C.~R.-C. acknowledges funding from the MathWorks Engineering Fellowship Fund by MathWorks Inc. R.~E.~C. acknowledges funding from Danmarks Grundforskningsfond (DNRF147) and Villum Fonden (8692). Y.S. acknowledges funding from the Swiss National Science Foundation (SNSF) through Project No. P2EZP2 188091.

% \printbibliography
\printbibliography[heading=subbibliography]

\setcounter{figure}{0}
\let\conjugatet\overline
\renewcommand{\thefigure}{S\arabic{figure}}

\newpage 
\section*{Towards 3D-Printed Inverse-Designed Metaoptics -- SUPPLEMENTARY INFORMATION}
\setcounter{section}{0}

\section{Impossibility of two-angle concentration within the ray-optics approximation}
In this section, we show that within the ray-optics approximation, reciprocal linear systems cannot realize concentration of two plane waves incident at two different angles to the same spot. We consider a general linear system, described by its ABCD scattering matrix $M$:
\begin{equation}
    M = \begin{bmatrix}
M_{11} & M_{12}\\
M_{21} & M_{22}.
\end{bmatrix}
\end{equation}
Reciprocity imposes symmetry on the scattering matrix (see for instance Ref. \cite{Haus1984WavesOptoelectronics}), which translates into : $\det(M) = M_{11} M_{22} - M_{12} M_{21} = 1$. The reference input plane is chosen to be the input facet of the system; the reference output plane is chosen to be the focal plane. Let us assume that this linear reciprocal system can focus  incident plane waves at two distinct angles (0 and $\theta$) in the same focal spot. The former incident angle translates into $M [ h, 0]^T = [0,\theta']^T$ for all $h$. This imposes the condition $M_{11} = 0$. The latter incident angle condition translates into $M [ h, \theta]^T = [0,\theta']^T$ which imposes $M_{12}=0$. This contradicts the reciprocity assumption. Therefore, a linear, reciprocal system \textit{cannot} realize concentration of two distinct angles within the ray-optics approximation. This fundamental limitation motivates our use of full-wave Maxwell equations-based model coupled with gradient-based inverse-design to identify optimized solutions to the concentrator problem, something that is not possible using analytical solution (unlike the thin-lens focusing problem). 

\section{Theoretical performance of plan-achromat multi-layer volumetric optics}

In Figure~1(d) of the main text, the simulated SR of multi-layer volumetric optics realizing multi-constrained functionalities (multi-angle, multi-wavelength, etc.) is shown. We used the following definition of the Strehl Ratio (SR) \cite{Hecht2016Optics}:
\begin{align}
    \mathrm{SR} = \frac{I(0)/\int_{A\rightarrow\infty} I(\mathbf{r})~d\mathbf{r}}{\mathrm{Airy}(0)/\int_{A\rightarrow\infty} \mathrm{Airy}(\mathbf{r})~d\mathbf{r}},
\end{align}
where the integration area $A$ over the focal plane is taken to infinity.

\section{Additional comparison: zero-shift design angles (theory and experiment)}

\begin{figure}
    \begin{center}
    \hspace*{-1.0in}
    \includegraphics[scale=0.85]{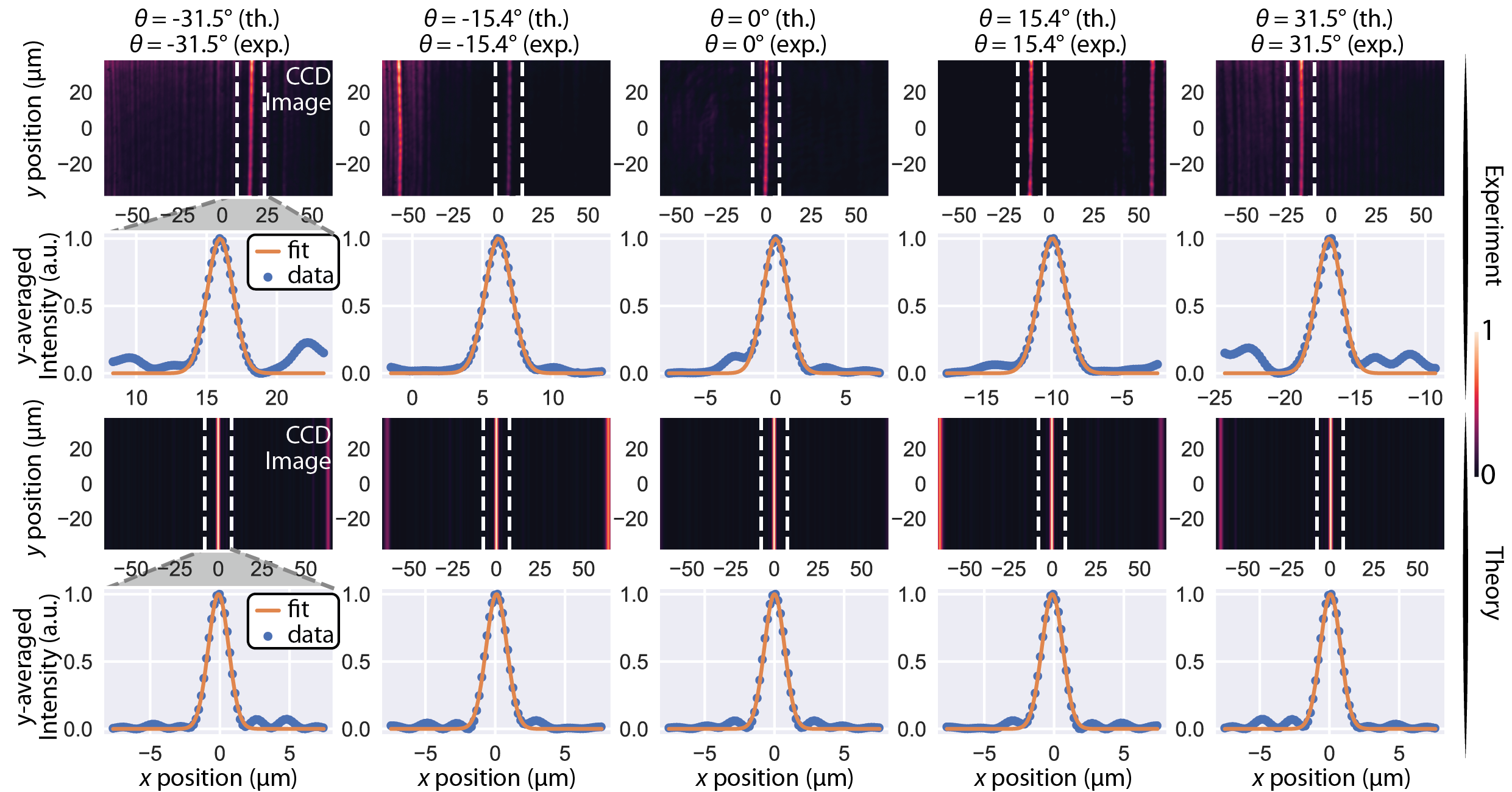}
    \caption{\textbf{Zero-shift design angles: comparison theory and experiment.} Columns correspond a given measured and simulated angle (whose values is indicated at the top of the column). The first two rows show experimental results. The last two rows show numerical simulations. In the rows showing the CDD images, the dashed white square corresponds to the area over which the signal is averaged to give the row below.}
    \label{fig:desang}
    \end{center}
\end{figure}

We commented in the main text that the measured zero-shift design angles were $\sim$ 12\% off, compared to the theoretical design. To confirm the zero-shift angle mismatch between experiments and numerical simulations, we plotted the measured CCD images and $y$-averaged focal lines at the design angles (and the theoretical prediction) in Figure~\ref{fig:desang}. As can be seen in Figure~\ref{fig:desang}, the designed zero-shift angles are $\theta_\text{th} \in \lbrace -31.5^\circ, -15.4^\circ,0,15.4^\circ, 31.5^\circ \rbrace$ but the measured focal lines at these design angles shift from the reference focal line by $\sim 15,6,0,-10,-17$~$\mu$m respectively.

As noted in the Discussion section of the main text, at least two factors could explain this mismatch: (1) refractive index inaccuracy for TPP IP-DIP; (2) fabrication inaccuracies and surface roughness.

\section{Efficiency: experiment and theory}

\begin{figure}
    \begin{center}
    \hspace*{-1.0in}
    \includegraphics[scale=0.85]{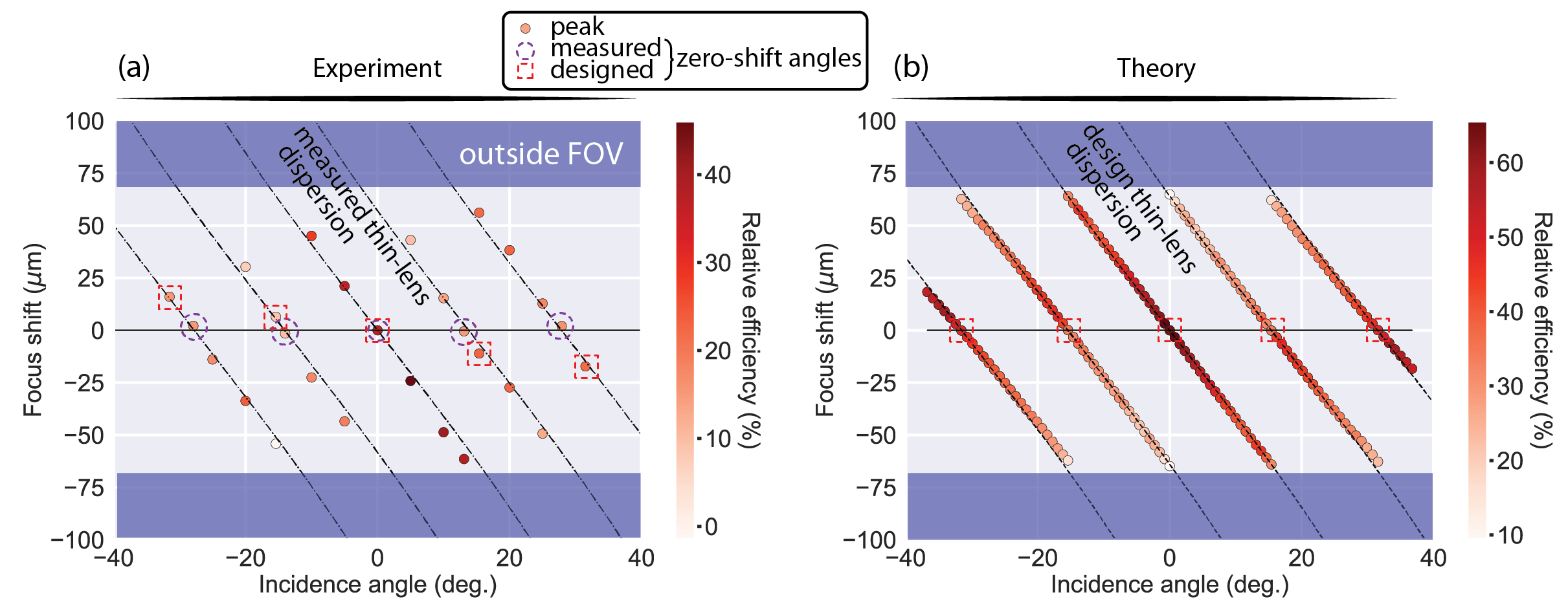}
    \caption{\textbf{Relative efficiency in Field of View.} (a) Experimental results. (b) Numerical results.}
    \label{fig:releff}
    \end{center}
\end{figure}

\begin{figure}
    \begin{center}
        \hspace*{-1.0in}
    \includegraphics[scale=0.85]{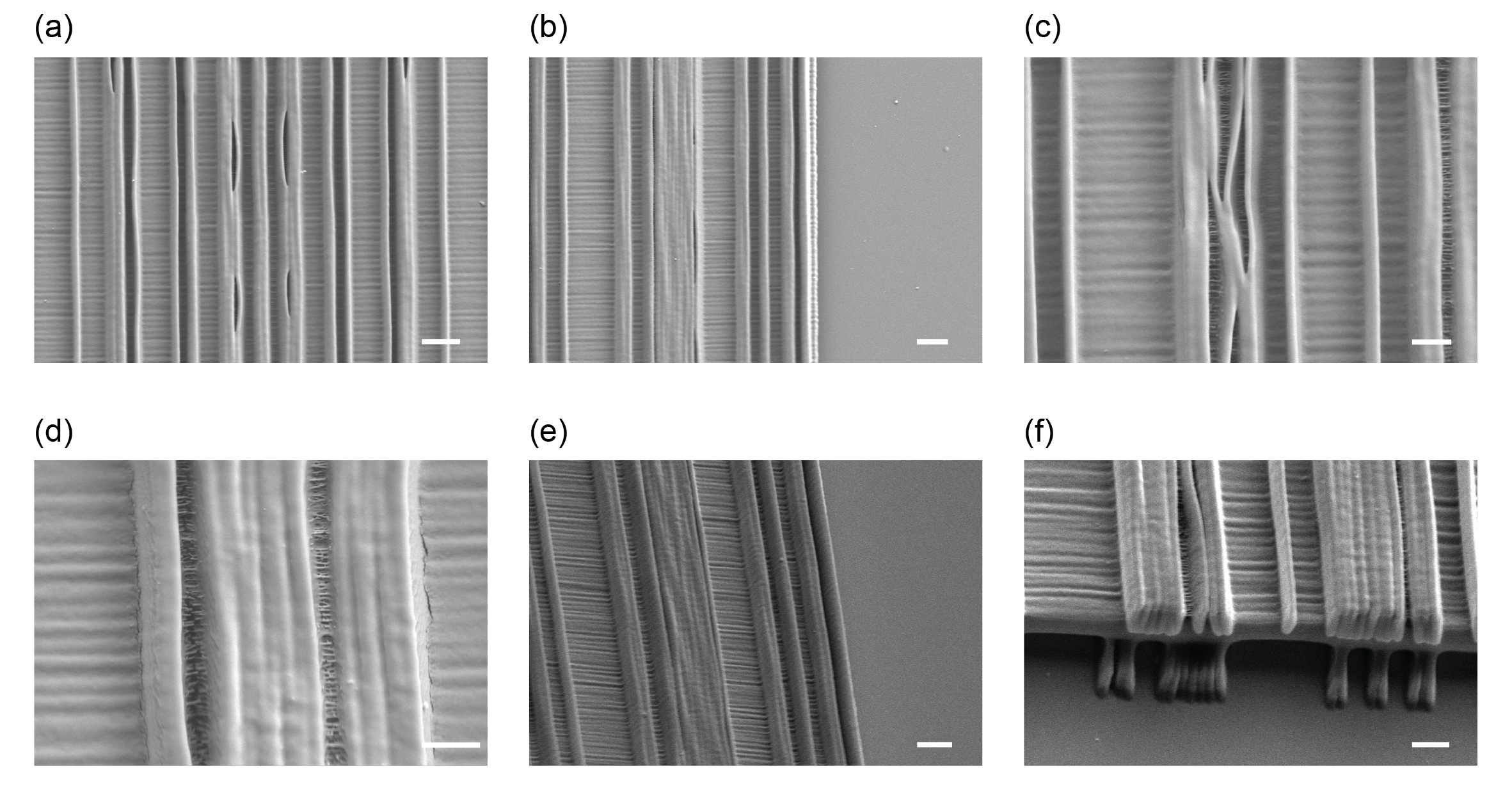}
    \caption{\textbf{Additional Scanning Electron Micrographs.} (a-d) Top views normal viewing angle. (e-f): Top views, tilted viewing angle (45$^\circ$). Scale bar (a-e): 2$\mu$m, 2$\mu$m, 1$\mu$m, 1$\mu$m, 2$\mu$m, 1$\mu$m.}
    \label{fig:moresem}
    \end{center}
\end{figure}

To further elaborate on the efficiency mismatch between experiment and theory, we plot the relative efficiency (normalized to the total measured power in the detector field of view) in Figure~\ref{fig:releff}. The mismatch is greatly reduced between experiment and theory when comparing the relative efficency: about a factor of 1.5 (as opposed to $\sim 5$ for absolute efficiency).

This supports the idea that the mechanism responsible for the decrease in efficiency acts by sending a significant amount of the scattered light outside of the detector field of view. This is typically what happens with surface scattering (whereby light is isotropically scattered at the surface by subwavelength defects). The additional SEM shown in Figure~\ref{fig:moresem} support this claim, since we identify a few areas where the extruded lines collapse on each other. This is a well-known problem when printing closely-spaced elongated structures. Typically, one can use critical point drying to reduce capillary forces acting while polymerizing the resist and washing away unpolymerized resist \cite{Christiansen2020FullwaveMetalenses}. 

\section{FWHM and focal distances}

The Full-Width at Half Maximum (FWHM) and focal distance of each focal line is measured experimentally and compared to the theoretical design.

\begin{figure}
    \begin{center}
    \includegraphics[scale=0.85]{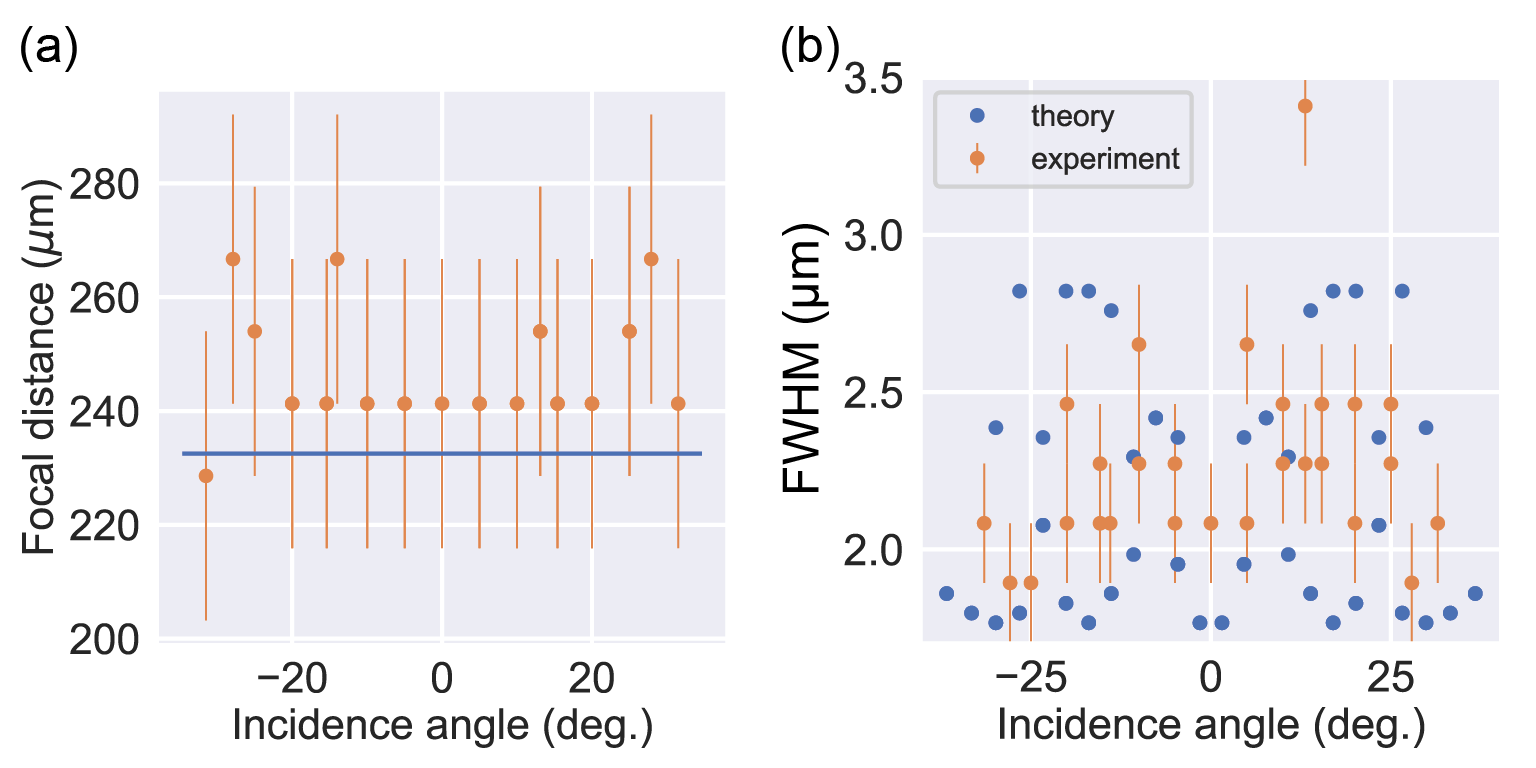}
    \caption{\textbf{FWHM and focal distance.} raelch: (a) Focal distance and (b) FWHM: numerical (blue dots) and experimental (orange squares) results with error bars.}
    \label{fig:fwhm}
    \end{center}
\end{figure}

The FWHM is obtained by fitting the $y$-averaged CCD image to a Gaussian function of the form $g(x) = C \exp \left( - \frac{-(x-x_\theta)^2}{2\sigma^2}\right)$ where $x_\theta$ is the $x$-position of the focal line (used to calculate the $x$-shift) and $\text{FWHM}=2{\sqrt  {2\ln 2}}\;\sigma \approx 2.355\;\sigma$. The relative uncertainty of the fit on $\sigma$ given the experimental data results in the error bars seen in Figure~\ref{fig:fwhm}(b)

The focal distance is obtained by moving the imaging assembly along the $z$ direction to measure the maximum unsaturated power of the focal line of interest. The minimum increment on the linear stage being 22.5 $\mu$m results on the corresponding error bars seen in Figure~\ref{fig:fwhm}(a). The measured focal distance corresponds to the designed one, within experimental inaccuracies. 

Data for both FWHM and focal distances as a function of the angle of incidence in air are shown in Figure~\ref{fig:fwhm}. 

\section{Calibration measurement}

\begin{figure}
    \begin{center}
    \includegraphics[scale=0.85]{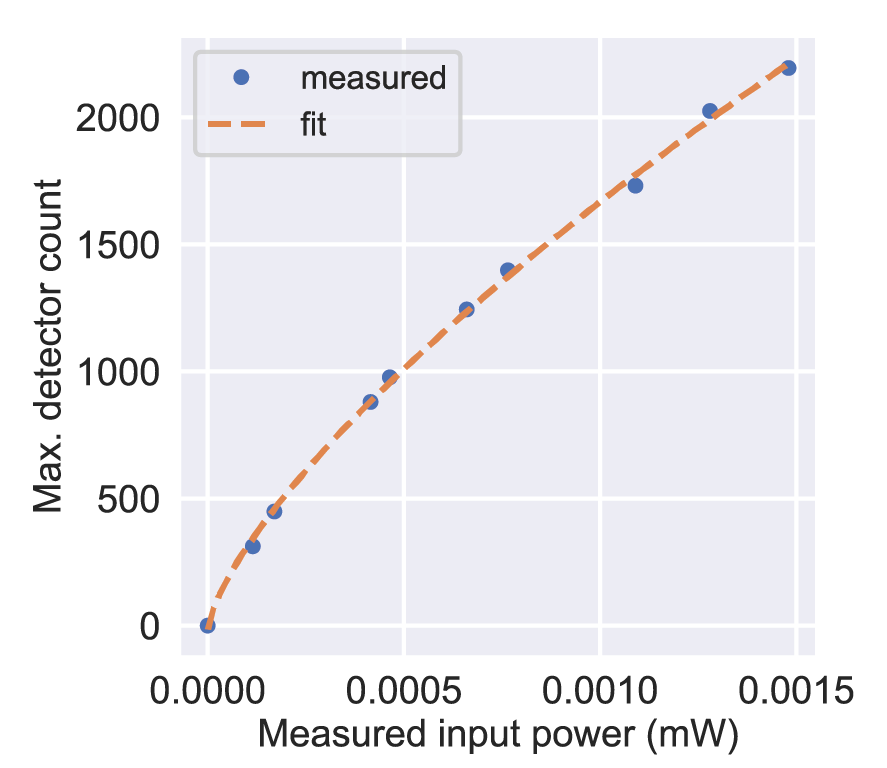}
    \caption{\textbf{Detector calibration.} Fit is given by $ y = 2.3\times10^5  \cdot x ^{0.71}$}
    \label{fig:calib}
    \end{center}
\end{figure}

We used an Electrophysics Infrared Imaging Camera (Model MicronViewer 7290A) to image the farfield in the focal plane. The detector's response can be calibrated by measuring the signal generated by a beam of known size and total power. The detector's response (assumed to be pixel-independent) is fitted to a power law shown in Figure~\ref{fig:calib}. The measured exponent of 0.71 matches the one given by the specification sheet within 2\%. This transfer function is used in our analysis to convert the measured signal to incident power per pixel. 

Additionally, we account for optical losses through the setup to calculate the absolute efficiency. The following components contribute to additional losses when the beam propagates through them: 100X objective, tube lens, sample substrate, collimation optics (since the incident power is measured before the collimation optics).

Lastly, we account for the mismatch between the incident beam size and the concentrator size. We measured the beam size with another infrared camera (Sony ICX445, sold by Edmund Optics) at the concentrator location on the optical axis. Assuming the beam is perfectly aligned with the concentrator, we can then estimate an upper bound of the incident power on the concentrator (per unit extruded length) $P_\text{in}$. In the event that the beam is not perfectly aligned with the center of the concentrator, the reported efficiencies are underestimated. Having a beam larger than the device allows us to achieve quasi-uniform illumination.

Using the detector's transfer function and the FWHM estimated from the Gaussian function fit, we can then estimate the total power in $\pm3~\text{FWHM}$ (per unit extruded length, since the power is averaged over the extruded direction). The resulting absolute efficiency is calculated as 
\begin{equation}
    \eta = \frac{P_{\pm3~\text{FWHM}}}{P_\text{in} ~ \eta_\text{opt}}.
\end{equation}

The relative efficiency shown in Figure~\ref{fig:releff} is instead normalized by the total power in the detector field of view (per unit extruded length) $P_\text{FOV}$:
\begin{equation}
    \eta = \frac{P_{\pm3~\text{FWHM}}}{P_\text{FOV} ~ \eta_\text{opt}},
\end{equation}
where $\eta_\text{opt}$ is the total optical loss through the various optical components in the setup.

\section{Optimized design and additional design considerations}

\begin{figure}
    \begin{center}
        \hspace*{-1.0in}
    \includegraphics[scale=0.85]{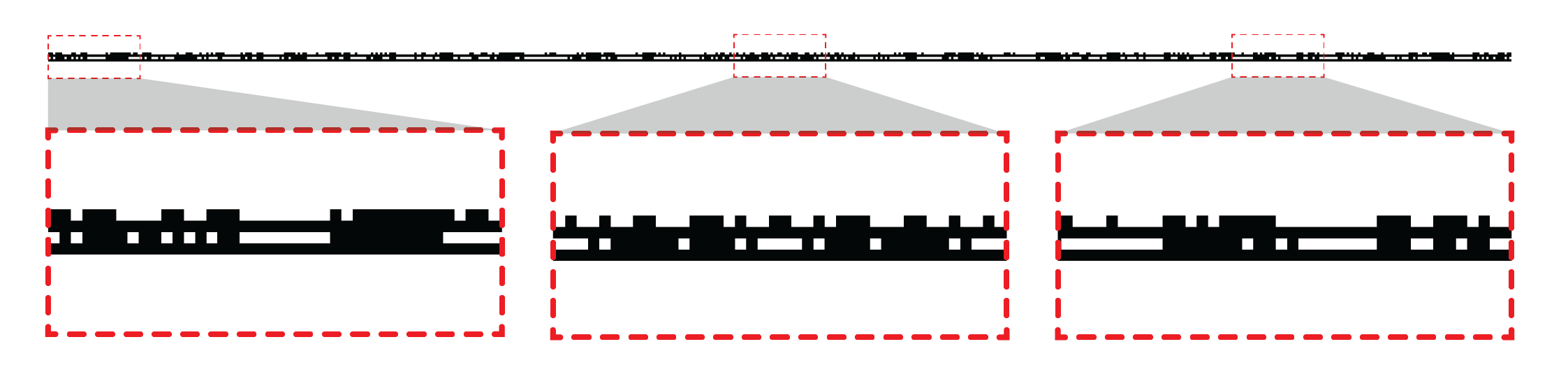}
    \caption{\textbf{Topology-optimized design.} Minimum feature size is 310 nm (horizontal) and 930 nm (vertical). The three insets show zoomed-in design regions.}
    \label{fig:designeps}
    \end{center}
\end{figure}

Topology optimization algorithms used in this paper have already been described in other papers \cite{Lin2018Topology-OptimizedMetaoptics, Lin2019OverlappingMetasurfaces, Lin2021ComputationalAberration, Christiansen2020FullwaveMetalenses}. Additional constraints were enforced to ensure the fabricability of the design (minimum feature size, total size, IP-DIP refractive index \cite{Gissibl2017RefractiveWriting}). The final design obtained through topology optimization is shown in Figure~\ref{fig:designeps}.

\begin{figure}
    \begin{center}
    \includegraphics[scale=0.85]{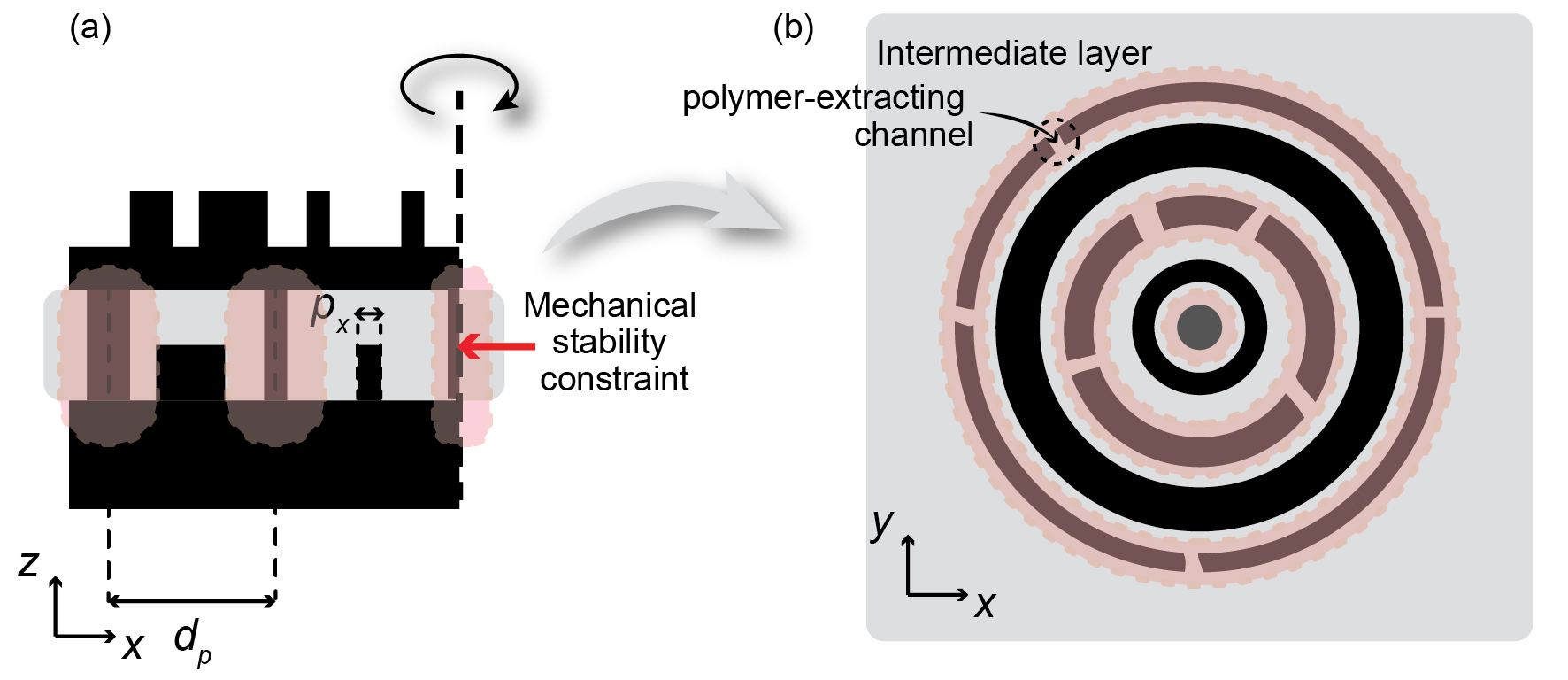}
    \caption{\textbf{Proposed design constraints for multi-layer full 3D cylindrically-symmetric devices.} (a) Radial cross-section. At a radial period of $d_p$, ridges are constrainted to connect the bottom and top layers via the patterned intermediate layer (while the other ridges allow an air layer of spacing). These extra tall ridges enable the mechanical stability of the full 3D structure with cylindrical symmetry, by supporting the top layer. (b) Top view of an intermediate layer. The existence of subwavelength polymer-extracting channels are added to the mechanical stability constraints in order to allow the polymer to evacuate through the channels during the development phase. This way the design does not present any closed cavity, despite the cylindrical symmetry and the multi-layering.}
    \label{fig:3ddesprop}
    \end{center}
\end{figure}

The proposed fabricable multi-layer designs in our framework are currently limited to extruded cylindrical lenses, which allow unpolymerized resist to be extracted. To utilize the full 3D potential of topology optimization, while satisfying fabrication constraints, we propose a few different approaches:
\begin{enumerate}
    \item Adding air layers in between each patterned-unpatterned layers, mechanically supported by taller pillars. To ensure extraction of unpolymerized resist, subwavelength holes would be randomly added along the ridges. The envisioned structures would rely on recently-proposed topology optimization techniques enforcing cylindrical symmetry, thus equivalent to a 2D optimization problem \cite{Christiansen2020FullwaveMetalenses}. A schematic of the concept is shown in Figure~\ref{fig:3ddesprop}.
    
    Going beyond a layered design with one of the following approaches:
    \item Enforcing mechanical stability and a "no-closed cavity" condition as additional constraints in the optimization formulation. Extensions of recent work on structural integrity \cite{Augenstein2020InverseIntegrity} could enable the realization of such free-form topologies with cylindrical symmetry. 
    \item Fabrication techniques allowing the printing of refractive index modulations with clustered nanoparticles in a solid matrix \cite{Oran20183DScaffolds}.
\end{enumerate}

\printbibliography
\end{document}